\documentclass[12pt]{iopart}

\usepackage{graphicx}
\usepackage{cite}
\usepackage{color}

\begin{document}
\bibliographystyle{iopart-num}

\title[The effect of ambipolar electric fields on the electron heating in CCRF plasmas]{The effect of ambipolar electric fields on the electron heating in capacitive RF plasmas}

\author{J. Schulze$^1$, Z. Donk\'o$^2$, A. Derzsi$^2$, I. Korolov$^2$, E. Schuengel$^1$}
\address{$^1$Department of Physics, West Virginia University, Morgantown, WV 26506, USA. \\ $^2$Institute for Solid State Physics and Optics, Wigner Research Centre for Physics,
Hungarian Academy of Sciences, 1121 Budapest, Konkoly Thege Mikl¬¨√≥\'os str. 29-33, Hungary.}
\ead{fjschulze@hotmail.com}

\begin{abstract}
We investigate the electron heating dynamics in electropositive argon and helium capacitively coupled RF discharges driven at 13.56 MHz by Particle in Cell simulations and by an analytical model. The model allows to calculate the electric field outside the electrode sheaths, space and time resolved within the RF period. Electrons are found to be heated by strong ambipolar electric fields outside the sheath during the phase of sheath expansion in addition to classical sheath expansion heating. By tracing individual electrons we also show that ionization is primarily caused by electrons that collide with the expanding sheath edge multiple times during one phase of sheath expansion due to backscattering towards the sheath by collisions. A synergistic combination of these different heating events during one phase of sheath expansion is required to accelerate an electron to energies above the threshold for ionization. The ambipolar electric field outside the sheath is found to be time modulated due to a time modulation of the electron mean energy caused by the presence of sheath expansion heating only during one half of the RF period at a given electrode. This time modulation results in more electron heating than cooling inside the region of high electric field outside the sheath on time average. If an electric field reversal is present during sheath collapse, this time modulation and, thus, the asymmetry between the phases of sheath expansion and collapse will be enhanced. We propose that the ambipolar electron heating should be included in models describing electron heating in capacitive RF plasmas.  
\end{abstract}

\pacs{52.80.Pi, 52.50.-b, 52.65.Rr, 52.25.Jm, 52.27.Aj} 
\submitto{Plasma Sources Science and Technology}

\maketitle

\section{Introduction}

Capacitively coupled radio frequency (CCRF) plasmas are frequently used for a great variety of technological applications ranging from Plasma Enhanced Chemical Vapor Deposition (PECVD) to plasma etching and medical applications such as sterilization or wound healing \cite{LiebermanBook,MakabeBook,ChabertBook}. Any effective optimization of these applications provides enormous societal benefits and must be based on a detailed scientific understanding of the plasma physics. This requires insights into the electron heating dynamics in CCRF plasmas to reveal how these discharges are generated and how they can be controlled for process optimization. 

Electropositive low pressure CCRF plasmas are typically operated in the $\alpha$-mode \cite{Belenguer1990,IOPConfBeams,Mahony}, while the $\gamma$- \cite{Belenguer1990,Schulze_Gamma} or the $\Omega$-mode \cite{Boeuf_Omega,Schulze_Omega,Hemke_Omega,Liu_Omega,SchungelDust,KillerDust} can be dominant in discharges at high pressure and/or high electronegativity. This $\alpha$-mode is characterized by dominant ionization/excitation maxima caused by  highly energetic electron beams generated during the phase of sheath expansion within the RF period at each electrode \cite{FTCBeams}. 

However, it is not clear how these electrons gain sufficient energy to cause ionization/excitation. The prevailing understanding of the electron heating mechanism is based on classical sheath expansion heating, i.e. the electrons adjacent to the sheath edge are accelerated by a single interaction with the expanding sheath into the plasma bulk. This is similar to Fermi heating \cite{HWModel2} and has been described by the Hard Wall Model \cite{HWModel} or other more elaborated models \cite{Godyak,Turner_Stoch,PrHeat1,PrHeat2,PrHeat3,Kaganovich1,Kaganovich2,Kaganovich3,Mussenbrock1,Mussenbrock2,Annaratone,LafleurHeat1,LafleurHeat2}. 

Here, we demonstrate that this understanding is not complete, via a modeling study of symmetric argon and helium plasmas driven at 13.56 MHz. In agreement with the Hard Wall Model each electron colliding with the expanding sheath gains twice the sheath expansion velocity. However, our Particle in Cell simulations complemented with Monte Carlo treatment of collision processes (PIC/MCC) show that the sheaths typically expand too slowly so that an electron cannot gain enough energy from a single interaction with the expanding sheath edge to cause ionization. While the mechanism of sheath expansion heating is present, we demonstrate that electrons are additionally heated by strong ambipolar electric fields around the position of maximum sheath width, where the normalized ion density gradient is high. This ambipolar electric field is required to couple the electron and ion fluxes locally \cite{LiebermanBook} and is located outside the sheath during most of the RF period. Therefore, it clearly represents a different heating mechanism based on a novel non-local kinetic effect: by tracing individual electrons in PIC/MCC simulations we demonstrate that electrons previously heated by interactions with the expanding sheath edge are additionally heated when propagating through this layer of ambipolar electric field. This effect is markedly different compared to previous works on diffusion phenomena at higher pressures. Moreover, we show that an electron can collide with the expanding sheath multiple times during one phase of sheath expansion, if it is scattered back towards the sheath by electron-neutral collisions.

In agreement with the Hard Wall Model, sheath expansion heating typically results in an increase of the electron velocity in the direction perpendicular to the electrodes of about $5-6 \times 10^5$ m/s per electron-sheath interaction. This corresponds to an energy increase of about 0.7 - 1 eV. In argon the ionization threshold is 15.6 eV corresponding to a total electron velocity of $v = \sqrt{v_x^2 + v_y^2 + v_z^2} \approx 2.3 \times 10^6$ m/s. For an electron temperature of about 3 eV classical sheath expansion heating alone cannot explain the observed ionization maximum during the phase of sheath expansion, since electrons in the high energy tail of the Electron Energy Distribution Function (EEDF) would not be able to reach high enough energies to cause the observed ionization rate. Moreover, experimentally observed electron beam velocities of about $2 \times 10^6$ m/s during the phase of sheath expansion cannot be explained by classical sheath expansion heating alone \cite{FTCBeams,Schulze_Coupling1,IEEEBeams}. We demonstrate that a synergistic combination of different heating events during one phase of sheath expansion, i.e. multiple interactions of one electron with the expanding sheath and/or acceleration by ambipolar electric fields outside the sheath, is required to accelerate an electron to energies exceeding the threshold for ionization. We find "ambipolar electron heating" to be essential to explain the generation of CCRF plasmas. 



The ambipolar field is found to be time modulated within the RF period, i.e. it is stronger during the phase of sheath expansion and weaker during sheath collapse. This represents an important asymmetry between the expansion and collapse phases that results in a different axial electric field profile during sheath expansion and collapse and finally in more heating than cooling on time average. Asymmetric electric field profiles during the phases of sheath expansion and collapse have been observed experimentally before \cite{IOPConfBeams,Schulze_Stoch,UCZ_EField}, but have never been explained until now. 


In argon the time modulation of the ambipolar electric field is found to be caused by a temporal modulation of the local mean electron energy. This is demonstrated by an analytical model and in agreement with models that identify pressure heating as the dominant collisionless electron heating mechanism \cite{Turner_Stoch,PrHeat1,PrHeat2,PrHeat3,LafleurHeat1,LafleurHeat2}. These models predict a net electron heating on time average only if the electron temperature in the sheath region is time modulated. These models, however, do not distinguish between sheath expansion and ambipolar heating, but include both mechanisms. Similarly sheath expansion and ambipolar electron heating have been discussed in the frame of fast PIC/MCC simulations before \cite{Schweigert,Ivanov}. The modulation of the electron mean energy is caused by the fact that highly energetic electrons previously heated by the expanding sheath cross the ambipolar field layer during the phase of sheath expansion, while low energetic electrons originating from the plasma bulk pass it during sheath collapse. 

An even stronger time modulation of the ambipolar field is observed in helium, where a strong electric field reversal \cite{LiebermanFieldRev,VenderFieldRev,TurnerFieldRev,UCZ_EField,SchulzeFieldRev} is present during sheath collapse. By an analytical model we demonstrate that this field reversal is caused by collisions with the neutral background gas and is required to draw enough electrons to each electrode during sheath collapse to compensate the positive ion flux at the electrode. This field reversal leads to an effective reduction of the overall electric field at the axial position of maximum ambipolar field during sheath collapse. This mechanism enhances the asymmetry between the sheath expansion and collapse phase and leads to even more electron heating on time average. 

We discuss this novel ambipolar electron heating mechanism for different pressures in argon and helium discharges operated at 13.56 MHz. The total heating rates of sheath expansion and ambipolar electron heating are compared and found to be of similar magnitude within one RF period.

The paper is structured in the following way: In sections 2 and 3, the PIC/MCC simulation and the analytical model to calculate the electric field are described briefly. In section 4, the results are presented. This part is divided into three subsections. First, we give an overview of the most important features of the electron heating mechanisms different from classical sheath expansion heating in both gases. In sections 4.2 and 4.3 we analyze peculiarities of argon and helium discharges separately in detail. Finally we draw conclusions in section 5.

\section{PIC/MCC simulation}

\begin{figure}[h!]
\begin{center}
\begin{tabular}{cc}
  \includegraphics[width=0.49\textwidth]{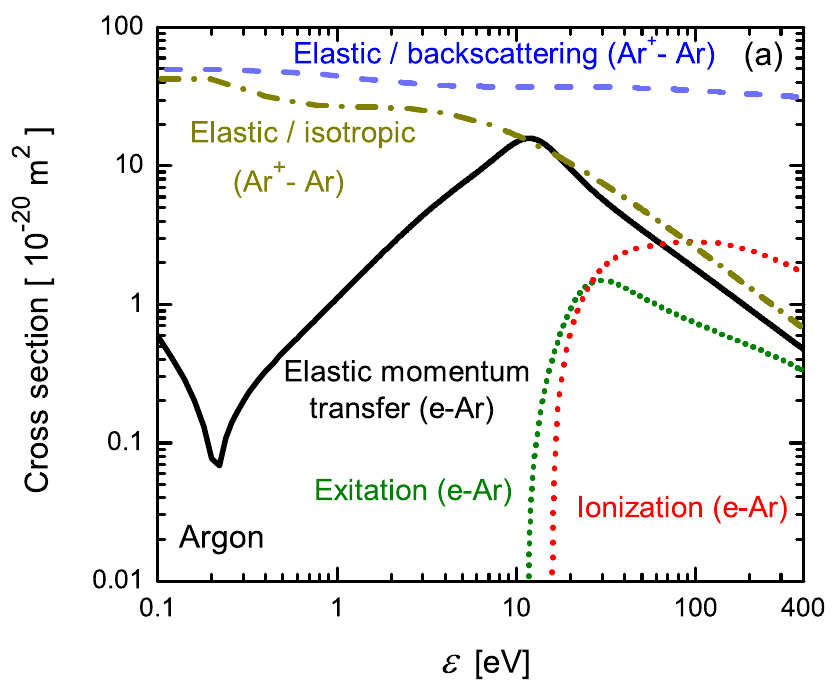}
 &
  \includegraphics[width=0.49\textwidth]{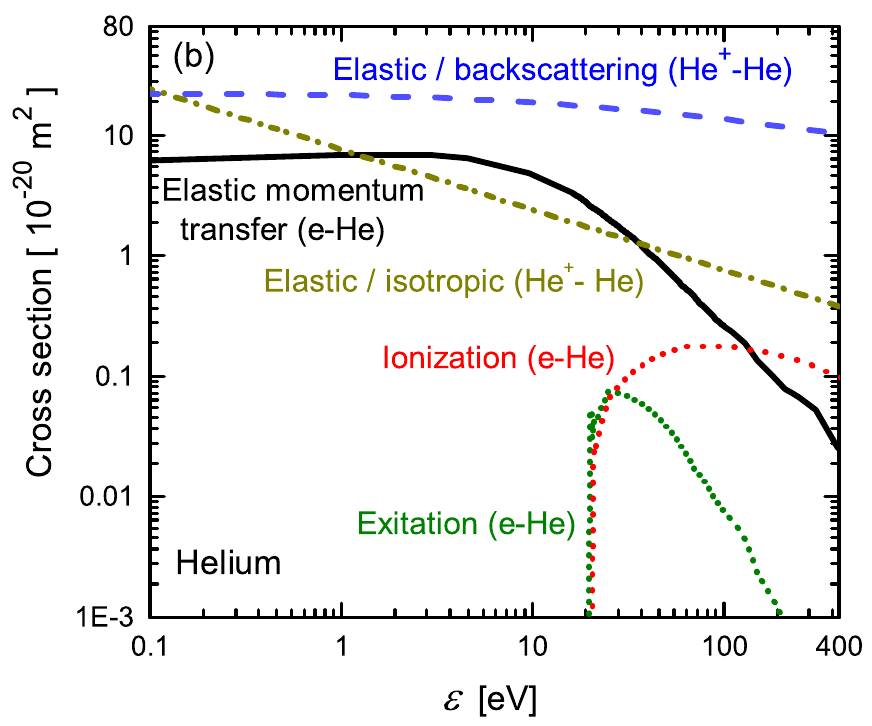}
 \\
\end{tabular}
\caption{Cross sections used in the frame of PIC/MCC simulation in argon (plot (a), \cite{sim1,sim3}) and helium (plot (b),\cite{sim4,sim2}).}
\label{cs}
\end{center}
\end{figure}

In our studies we use a one-dimensional (1d3v) bounded plasma PIC simulation code, complemented with Monte Carlo treatment of collision processes (PIC/MCC \cite{Birdsall,DP2006,PIC}). The electrodes are assumed to be infinite, planar and parallel, separated by a gap $L$. The $x$-direction is perpendicular to the electrodes. In our implementation of the PIC simulation, one of the electrodes (the ``bottom" electrode at $x=0$) is driven by the following (single frequency) voltage waveform

\begin{equation}
\phi(t) = \phi_0 \cos{(2 \pi f t)},
\label{Voltage}
\end{equation}
while the ``top" electrode (at $x=L$) is grounded.

The cross sections for electron-neutral and ion-neutral collisions - taken from \cite{sim1,sim3} and \cite{sim4,sim2} for argon and helium, respectively - are shown in figure \ref{cs}. The scattering angles after each collision are determined based on a Monte Carlo procedure described in detail in \cite{PIC2}. Electrons are reflected from the electrode surfaces with a probability of 0.2 \cite{Kollath_Alpha} and the secondary electron emission coefficient of positive ions at the electrodes is taken to be $\gamma = 0.1$. From the trajectories of the particles followed in the PIC simulation, as well as from the collision events we derive the spatio-temporal distributions of several discharge characteristics (e.g.  densities, electron heating, ionization rates, etc.). The total number of superparticles (electrons + ions) in the code is $\approx 1 \times 10^6$.

\section{Analytical model to calculate the electric field}

We use a fluid model to calculate the electric field outside the sheaths. This model was developed and described in detail in \cite{SchulzeFieldRev}. It is based on the electron momentum balance and continuity equations. Combining these two equations the electric field, {\it E}, is found to be the sum of four components, i.e.:

\begin{equation}
E = \sum_{i = 1}^{4}{E_i} 
\label{EField}
\end{equation}

with

\begin{equation}
E_1 = \frac{m_e}{n_e e^2} \frac{\partial j_e}{\partial t}
\label{term1}
\end{equation}

\begin{equation}
E_2 = \frac{\Pi}{ne}
\label{term2}
\end{equation}

\begin{equation}
E_3 = \frac{m_e}{n_e^3 e^3} \frac{\partial n_e}{\partial x} j_e^2
\label{term3}
\end{equation}

\begin{equation}
E_4 = E_{amb} = -\frac{k T_e}{e} \frac{1}{n_e} \frac{\partial n_e}{\partial x}.
\label{term4}
\end{equation}



Here, $m_e$ is the electron mass, $e$ is the elementary charge, $n_e$ is the electron density, $j_e$ is the electron conduction current density, $k$ is the Boltzmann constant, and $T_e$ is the electron temperature. In equation (\ref{term2}), $\Pi$ is the electron momentum loss per volume and time, which is computed directly in the simulation. In this way $E_2$ is determined in a more accurate way compared to its classical expression of $E_2^{class} = \frac{m_e \nu_c}{n_e e^2} j_e$, which assumes that the electrons' momenta are completely lost in each collision (here $\nu_c$ is the total electron collision frequency).
$j_e$, $n_e$, and the mean electron energy, $\langle \varepsilon \rangle$, needed in the calculation of the electric field, are taken from the simulation, as a function of position between the electrodes and time within the RF period.

Each of the four terms, (\ref{term1}) - (\ref{term4}), contributing to the electric field, represents a distinct physical mechanism. Applying this model allows to separate their individual contributions and, therefore, to identify the relevance of different physical mechanisms for the generation of the electric field obtained from the simulation at distinct positions and times. Equations (\ref{term1}) and (\ref{term3}) represent electron inertia, equation (\ref{term2}) represents collisions of electrons with the neutral background gas, and equation (\ref{term4}) is the ambipolar electric field. The sign of the fourth term is different from the signs of the first three terms. This means that the ambipolar electric field will accelerate electrons into the plasma bulk, if the plasma density decays monotonically from the discharge center towards the electrodes, and the first three terms can cause an electric field that accelerates electrons towards the electrodes. An electric field that accelerates electrons towards the electrodes is called a reversed field here. We note that the expression (\ref{term4}) for the ambipolar electric field assumes a Maxwellian distribution function; accordingly we approximate the electron temperature from the mean energy as $k T_e = (2/3) \langle \varepsilon \rangle$. 

\section{Results}


All computations have been carried out for an electrode gap of $L = 5$ cm, and for a fixed frequency, $f = 13.56$ MHz, and driving voltage amplitude, $\phi_0 = 400$ V. For argon, we discuss discharges operated at neutral gas pressures of 2 Pa and 20 Pa to probe relatively collisionless and collisional regimes. In helium, we focus on a pressure of 120 Pa, due to similar sheath width and expansion velocities compared to the argon case at 20 Pa as well as the presence of a strong electric field reversal during sheath collapse, that is not observed in case of the argon simulations. The field reversal is found to affect the dynamics of sheath expansion and ambipolar electron heating significantly during sheath collapse. In this way the field reversal affects the time averaged electron heating rate.

\subsection{Electron Heating Mechanisms}

 \begin{figure}[ht!]
 \begin{center}
 \includegraphics[width=0.92\textwidth]{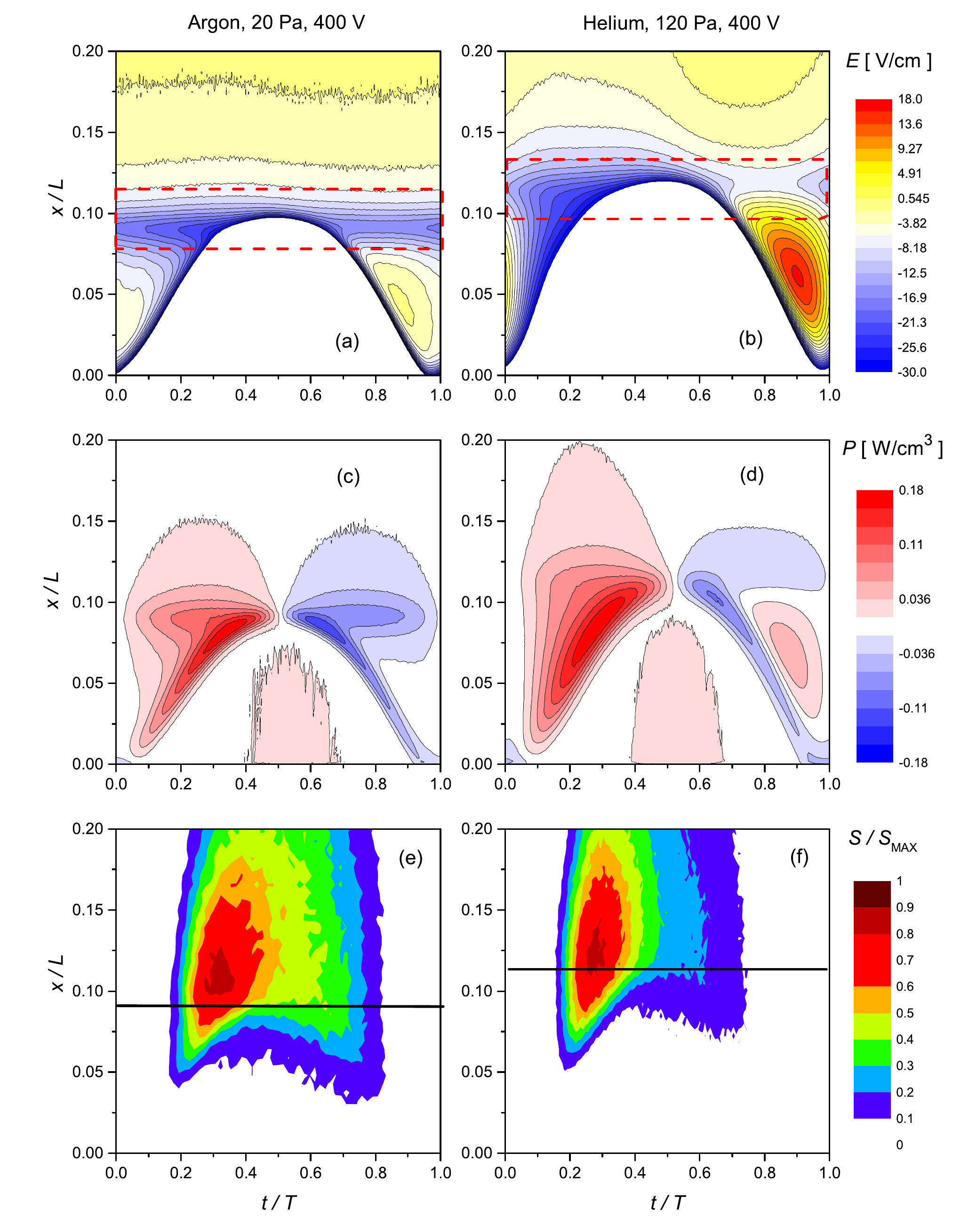} 
 \caption{PIC simulation results for the spatio-temporal distributions of the electric field (top row), electron heating rate, $P$ (middle row), and normalized ionization rate, $S/S_{MAX}$ (bottom row). Discharge conditions: argon, 20 Pa, 5 cm gap, $\phi_0$ = 400 V (left column) and helium, 120 Pa, 5 cm gap, $\phi_0$ = 400 V (right column). Only the spatial region close to the bottom (powered) electrode is shown. The horizontal axis shows the time normalized by the duration of one RF period, $T$. $S_{\rm MAX}$ = 2.2$ \times 10^{21}$ m$^{-3}$s$^{-1}$ and 6.0$ \times 10^{21}$ m$^{-3}$s$^{-1}$, for (e) and (f), respectively.}
 \label{overview}
 \end{center}
 \end{figure}

\begin{figure}[h!]
\begin{center}
\begin{tabular}{c}
  \includegraphics[width=1\textwidth]{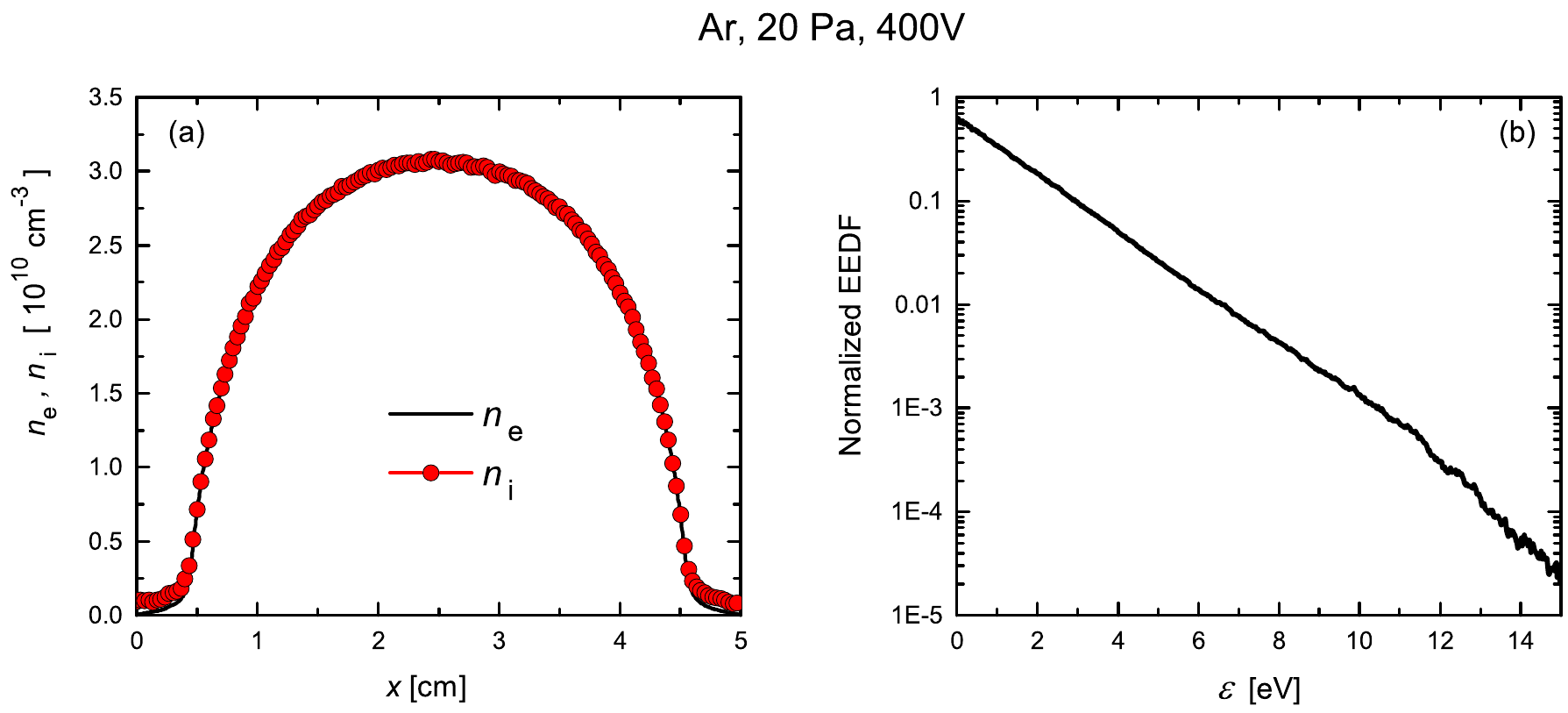}
\\
  \includegraphics[width=1\textwidth]{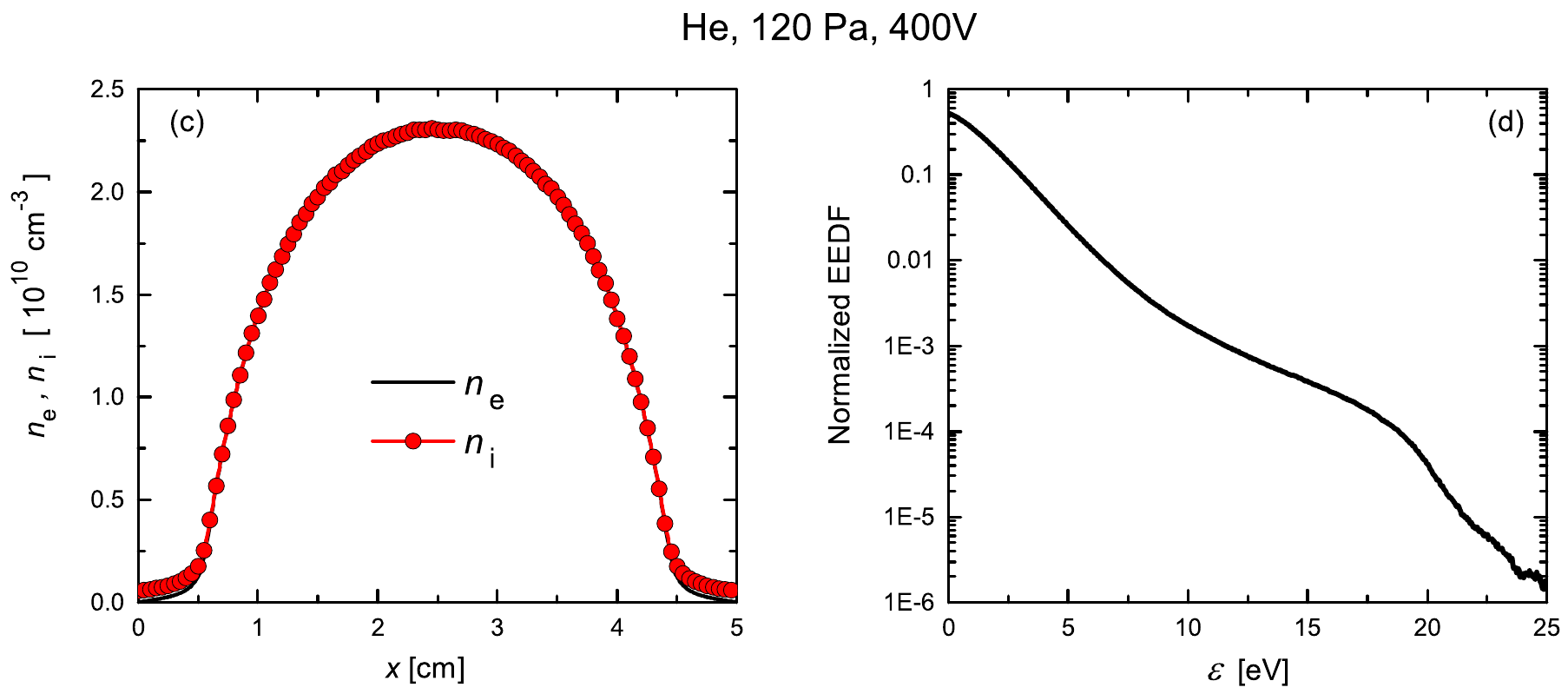}
 \\
\end{tabular}
\caption{Time averaged profiles of the electron and ion density as well as the time averaged normalized EEDF in the discharge center in argon at 20 Pa, 5 cm gap, and $\phi_0$ = 400 V [plots (a) and (b)] and in helium at 120 Pa, 5 cm gap, and $\phi_0$ = 400 V [plots (c) and (d)] obtained from the simulation.}
\label{DensEEDF1}
\end{center}
\end{figure}

The top row of figure \ref{overview} shows the spatio-temporal distribution of the electric field close to the bottom (powered) electrode in an argon discharge operated at 20 Pa [fig. \ref{overview} (a)] and in a helium discharge operated at 120 Pa [fig. \ref{overview} (b)]. These conditions are chosen for comparison, since they result in similar maximum sheath widths and sheath expansion velocities in these two gases, but the presence of a strong electric field reversal during sheath collapse only in helium. This allows to clarify the effect of the presence of the field reversal on different heating mechanisms. The horizontal axes in figure \ref{overview} cover one RF period at 13.56 MHz ($T \approx 74$ ns). The sheaths are clearly visible as white regions in these plots.
The time averaged electron and ion density profiles as well as the time averaged and normalized EEDF in the discharge center are shown in figure \ref{DensEEDF1} for both cases. In helium the plasma density in the bulk is lower compared to the argon case and the EEDF is non-Maxwellian. 

In most models of CCRF discharges the sheath edge is assumed to be a Hard Wall, where the electric field drops steeply from a high value inside the sheath to zero outside the sheath. Figures \ref{overview} (a) and (b) show that this is not true and is not a good approximation to model the electron heating dynamics. 

In argon [fig. \ref{overview} (a)], there is a horizontal zone of high electric field located at the position of maximum sheath width, $x/L \approx 0.09$, indicated by the dashed rectangle. This region is located outside the sheath during most of the RF period. It is located inside the sheath, only when the sheath is fully expanded. In this region, electric fields of up 20 V/cm and a voltage drop of about 4 V are found outside the sheath. Apart from the specific values, this is a general phenomenon observed at different conditions. While the highest electric fields are found inside the sheath, there are only a few secondary electrons that are accelerated by these strong fields. However, many electrons are present and accelerated at the position of high electric field outside the sheath. Thus, these electric fields -- not included in the Hard Wall Model -- contribute significantly to the overall electron heating. In sections 4.2 and 4.3 we will demonstrate that this field is an ambipolar electric field that must not be neglected, when analyzing the electron heating dynamics in CCRF plasmas. In helium [fig. \ref{overview} (b)], a similar region of high electric field outside the sheath is observed around $x/L \approx 0.11$. In addition, there is also a strong field reversal during sheath collapse in helium \cite{LiebermanFieldRev,VenderFieldRev,TurnerFieldRev,UCZ_EField,SchulzeFieldRev}. Similar ambipolar electric field structures and field reversal effects have been observed by Braginsky et al. in CCRF discharges driven at a lower driving frequency of 1.76 MHz \cite{Braginsky}.

The horizontal layer of high electric field outside the sheath results in strong electron heating during the first half of the RF period, when the sheath expands, and cooling during the second half, when the sheath collapses [figs. \ref{overview} (c) and (d)]. This heating and cooling along the zone of high electric field outside the sheath is observed in addition to the known sheath expansion heating and cooling that is located directly at the sheath edge. Thus, the electron heating and cooling inside the horizontal layer of high electric field represents a different distinct heating/cooling mechanism. There is also weak additional electron heating inside the sheaths due to the acceleration of ion induced secondary electrons around the time of maximum sheath expansion \cite{Belenguer1990,Schulze_Gamma}. In helium, the electric field reversal during sheath collapse yields additional heating during the second half of the RF period.

A maximum of the ionization is observed during sheath expansion [figs. \ref{overview} (e) and (f)] in both gases. This is well-known, but has been purely attributed to classical sheath expansion heating until now, i.e. a single interaction of individual electrons with the expanding sheath. Here, we observe that this is not correct. These ionization maxima are not observed at the sheath edge, but shortly above the horizontal layer of high electric field indicated by the horizontal black lines in plots (e) and (f). We will demonstrate that these maxima are caused by electrons accelerated by different heating events, i.e. multiple interactions with the expanding sheath within one RF period and electron heating by ambipolar electric fields outside the sheath.

The electric field outside the sheath is time dependent [figs. \ref{overview} (a) and (b)], i.e. it is stronger during the phase of sheath expansion compared to the collapse. This represents an important asymmetry between the two halves of one RF period at one electrode and results in more heating than cooling on time average. It also results in different axial profiles of the electric field during sheath expansion and collapse. Similar asymmetries have been observed experimentally by laser electric field measurements before \cite{IOPConfBeams,Schulze_Stoch,UCZ_EField}, but have neither been explained nor considered in models to describe electron heating in CCRF plasmas yet. This asymmetry is more pronounced in helium compared to argon due to the presence of the field reversal during sheath collapse, which effectively reduces the absolute value of the electric field in the horizontal layer located around the position of maximum sheath width during sheath collapse. It also leads to a high electric field adjacent to the collapsing sheath edge that accelerates electrons towards the electrode. In sections 4.2 and 4.3 we will demonstrate that this asymmetry of the electric field outside the sheath between sheath expansion and collapse is caused by the time modulation of the electron mean energy and the field reversal (helium).

\newpage
\subsection{Argon plasmas}

 \begin{figure}[ht!]
 \begin{center}
 \includegraphics[width=0.54\textwidth]{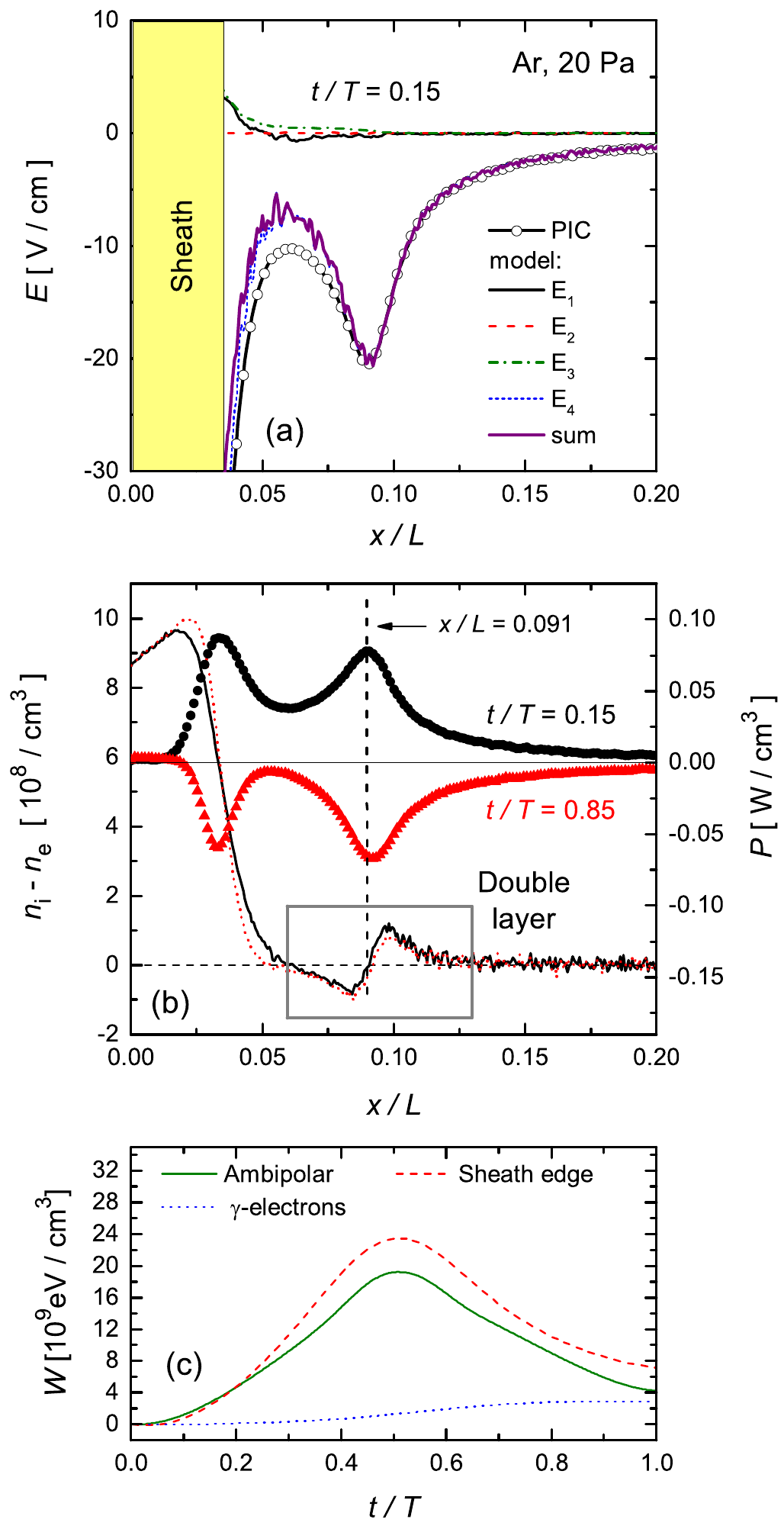} 
 \caption{(a) Axial electric field profile in the vicinity of the powered electrode during the phase of sheath expansion ($t/T = 0.15$) resulting from the simulation (PIC) as well as profiles of each term in equation (\ref{EField}) separately, and the sum of all terms. (b) Spatial profile of the net density, $n_i - n_e$ (lines), and electron heating rate, $P$ (symbols), during sheath expansion ($t/T = 0.15$, solid black line) and collapse ($t/T = 0.85$, dashed red line). (c) Energy transferred to electrons from the beginning of the RF period, $W$, as a function of $t/T$ within one RF period in three different spatial regions: ambipolar field ($x/L=0.091$, green solid line), at the sheath edge (red dashed line), and inside the sheath (blue dotted line). Discharge conditions: argon, 20 Pa, 5 cm gap, $\phi_0 = 400$ V.}
 \label{Efield_Analysis_Ar}
 \end{center}
 \end{figure}

Panel (a) of figure \ref{Efield_Analysis_Ar} shows the axial electric field profile close to the powered electrode during the phase of sheath expansion ($t/T = 0.15$) resulting from the simulation in argon at 20 Pa, as well as profiles of each term in equation (\ref{EField}) separately, and the sum of all terms.  This time was chosen for the analysis, since the position of the sheath edge and that of the maximum ambipolar field can clearly be distinguished. At later times these two axial positions move closer to each other and finally coalesce at the time of maximum sheath expansion (see fig. \ref{overview}). The time of $t/T \approx 0.85$ during sheath collapse, which is analyzed in the middle plot of figure \ref{Efield_Analysis_Ar} was chosen for symmetry reasons ($0.85 = 1 - 0.15$). At $t/T = 0.15$ the electric field profile obtained from the simulation is reproduced well by $E_4$ based on equation (\ref{term4}). $E_1$, $E_2$, and $E_3$ are negligible under these conditions. Thus, the local extremum of the electric field outside the sheath at $x/L \approx 0.09$ is an {\it ambipolar electric field} caused by the steep local gradient of the plasma density profile, and couples the motion of electrons and positive ions.

Figure \ref{Efield_Analysis_Ar}(b) shows that the presence of this ambipolar electric field is accompanied by a double layer of different space charges around $x/L \approx 0.09$. According to Maxwell's equations this double layer is a consequence of the local extremum of the axial electric field profile. The space charge is positive on the bulk side of the horizontal layer of ambipolar electric field and negative on the sheath side. This is caused by the fact that electrons move faster towards the electrodes than the positive ions, but are accelerated towards the bulk by the ambipolar field to couple the ion and electron motions. Figure \ref{Efield_Analysis_Ar}(b) also shows that the electron heating rate, $P$, is strong at two different axial positions close to the powered electrode during the phase of sheath expansion ($t/T = 0.15$): $P$ is high at the sheath edge due to sheath expansion heating and high within the region of high ambipolar field at $x/L \approx 0.09$. Both maxima are similar, i.e. {\it both electron heating mechanisms are of similar relevance} at this time within the RF period. Two mechanisms of electron cooling are observed at $t/T \approx 0.85$: electrons are cooled, when they enter the region of high ambipolar electric field ($x/L \approx 0.09$) and when they enter the collapsing sheath ($x/L \approx 0.03$). The electron cooling due to both mechanisms is lower compared to the respective heating during sheath expansion at $t/T = 0.15$.

 \begin{figure}[ht!]
 \begin{center}
 \includegraphics[width=0.6\textwidth]{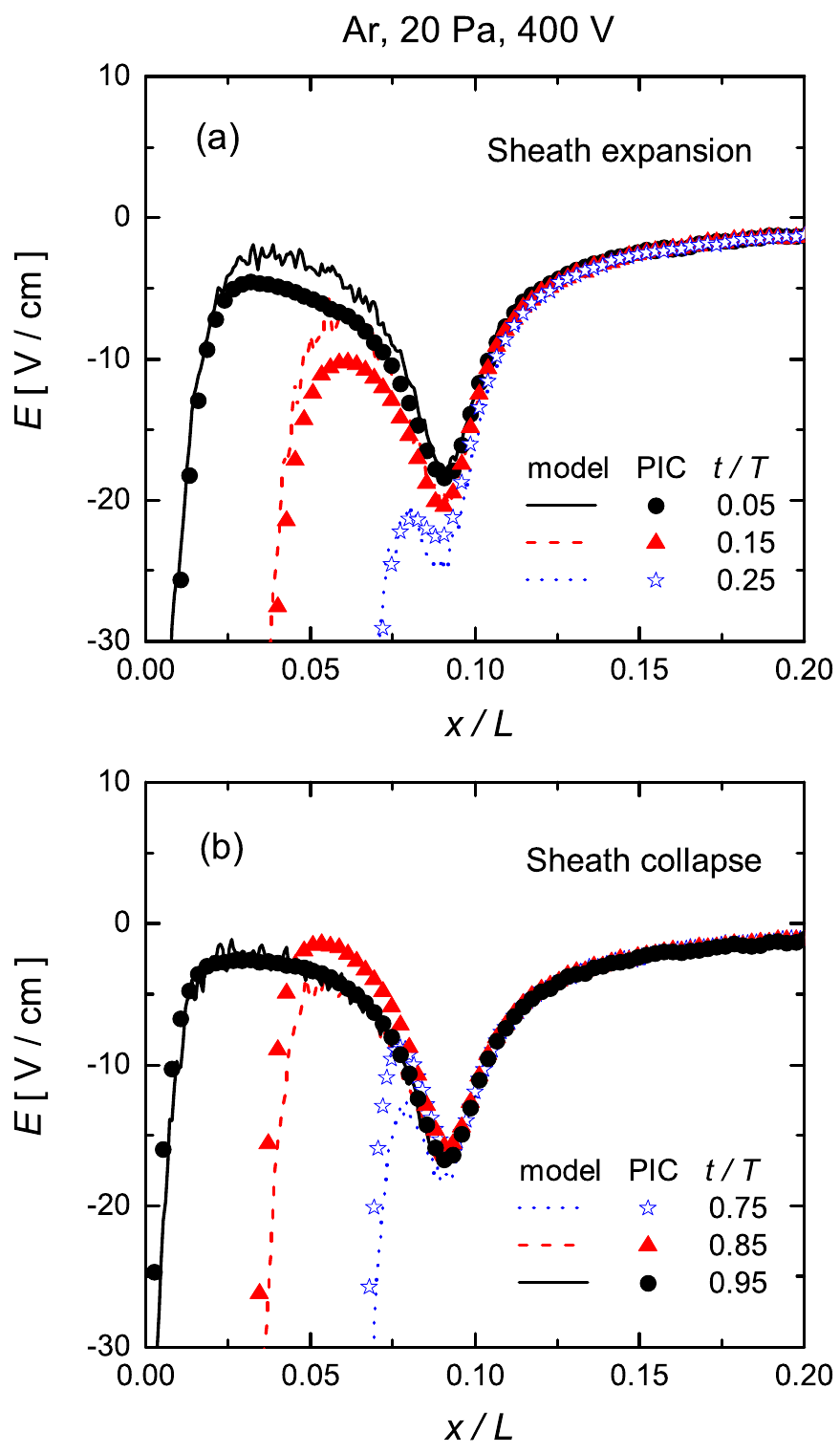} 
 \caption{Axial profiles of the electric field close to the powered electrode obtained from the simulation and the analytical model [equation (\ref{term4})] at 3 distinct times during sheath expansion (a) and collapse (b). Discharge conditions: argon, 20 Pa, 5 cm electrode gap, $\phi_0 = 400$ V.}
 \label{Field_Mod_Ar}
 \end{center}
 \end{figure}

Figure \ref{Efield_Analysis_Ar}(c) shows the energy transferred to electrons since the beginning of the RF period, $W (t/T) = \int_0^{t/T}{P(t'/T) d(t'/T)}$, as a function of $t/T$ within one RF period in three different spatial regions, i.e. the region of high ambipolar electric field ($x/L=0.091$, green solid line), at the sheath edge ($s(t)-0.1$ mm $\leq x \leq s(t)+0.1$ mm, red dashed line), and inside the sheath ($x\leq s(t)-1$ mm, blue dotted line). Here, $s(t)$ is the position of the sheath edge as a function of time \cite{Brinkmann}. The solid green line corresponds to the ambipolar electron heating, the dashed red line corresponds to sheath expansion heating, and the dotted blue line corresponds to the electron heating related to secondary electrons inside the sheaths. Under these conditions the heating due to secondary electrons yields the lowest contribution to the total energy transfer. It increases monotonically as a function of time, since secondary electrons are not cooled, but only accelerated inside the sheath. The strongest contribution is related to sheath expansion heating. The ambipolar electron heating is roughly half as strong as the sheath expansion heating after one full RF period, but clearly contributes significantly to the total energy transfer within one RF period and must not be neglected. It shows a similar time dependence compared to sheath expansion heating, i.e. there is heating during the phase of sheath expansion and cooling during sheath collapse, but the cooling is weaker than the heating due to a time dependence of the electric field outside the sheath.

This time modulation is illustrated in figure \ref{Field_Mod_Ar}, where the axial electric field profiles obtained from the simulation and the analytical model are shown close to the powered electrode at 3 distinct times during sheath expansion [plot (a)] and collapse [plot (b)], respectively. The electric field outside the sheath is clearly stronger during the phase of sheath expansion compared to the phase of sheath collapse.

 \begin{figure}[ht!]
 \begin{center}
 \includegraphics[width=0.65\textwidth]{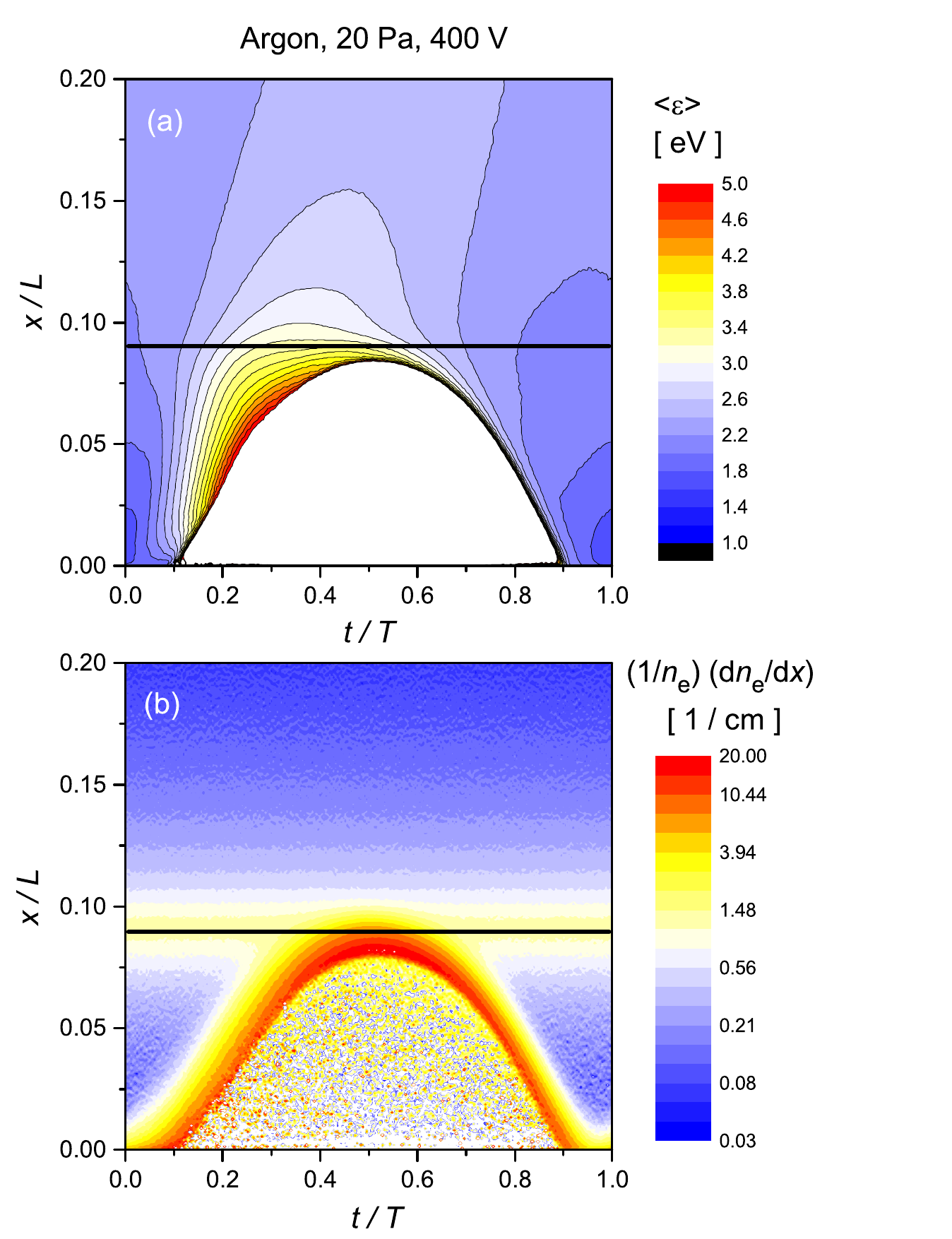} 
 \caption{Spatio-temporal plots of the electron mean energy, $<\varepsilon>$, (a) and the normalized density gradient in the fourth term of equation (\ref{EField}), $(1/n_e) (dn_e/dx)$, (b) close to the powered electrode. These results are obtained from the simulation. Discharge conditions: argon, 20 Pa, 5 cm gap, $\phi_0$ = 400 V. (The noise within the sheath region in (b) originates from the low density of electrons.)}
 \label{Time_Dep_Ar}
 \end{center}
 \end{figure}

The reason why the ambipolar electric field is time dependent is crucial to understand why there is a net positive electron heating on time average due to the ambipolar field. It can be understood based on the analytical model. The ambipolar electric field is described by equation (\ref{term4}) and can be time modulated due to a time modulation of the electron temperature, i.e. the electron mean energy $<\varepsilon> = \frac{3}{2} k T_e$, and/or a time modulation of the normalized density gradient, $1/n_e \cdot \partial n_e / \partial x$. Figure \ref{Time_Dep_Ar} shows spatio-temporal plots of both factors. Around the position of maximum ambipolar electric field ($x/L \approx 0.09$) only the electron mean energy is time modulated as long as this position is located outside the sheath. When it is located inside the sheath a time modulation of the normalized density gradient is observed, since the electron density decreases. Based on these results the reason for the time modulation of the ambipolar electric field outside the sheath is clearly the time modulation of the electron mean energy. The time modulation of $<\varepsilon>$ in turn is caused by the sheath expansion heating present only during the first half of the RF period. Electrons heated by the expanding sheath below the layer of high ambipolar field pass this layer with high mean energies. This does not happen during sheath collapse, when cold electrons from the bulk enter the ambipolar layer.

 \begin{figure}[ht!]
 \begin{center}
 \includegraphics[width=1\textwidth]{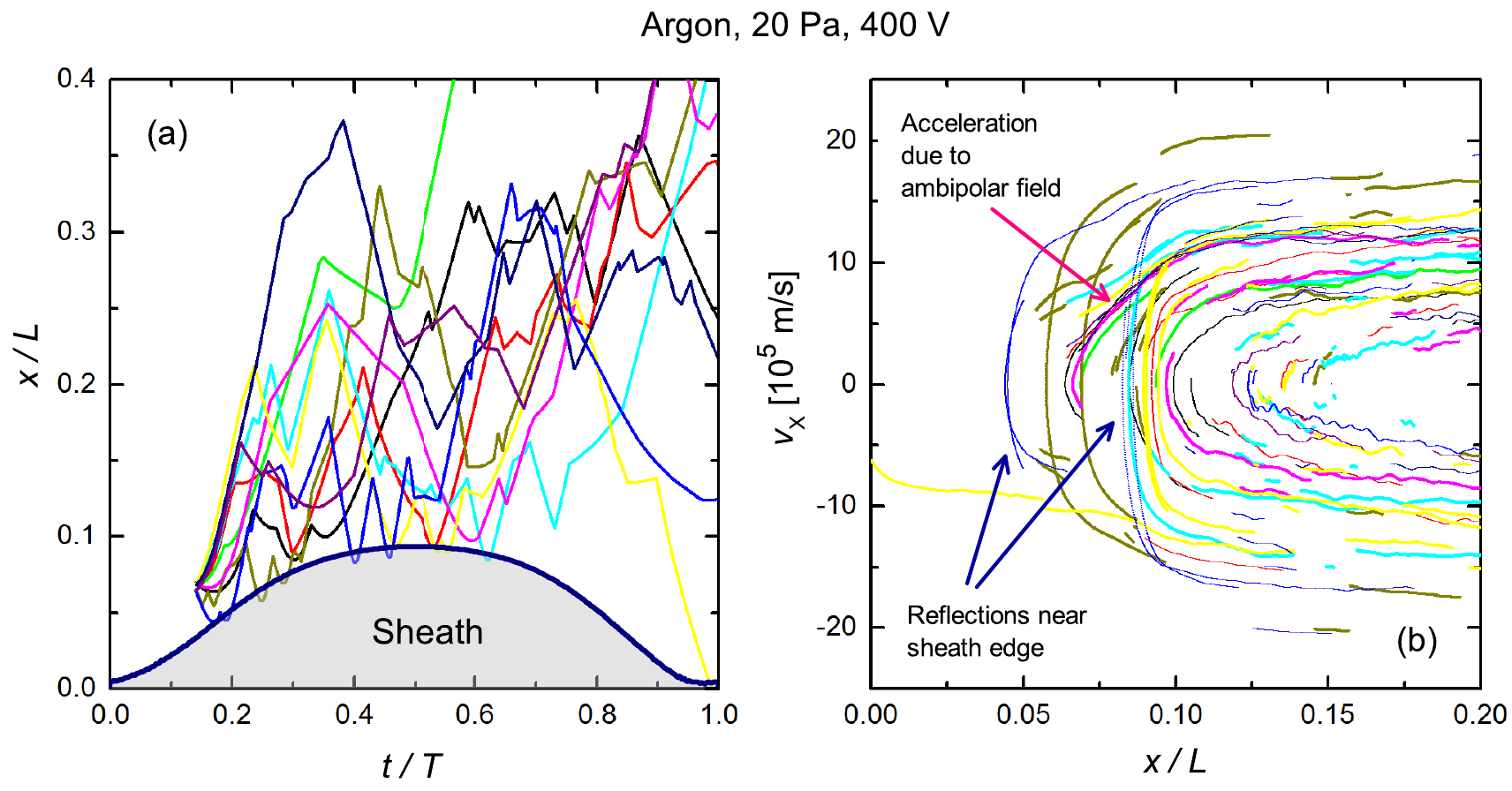} 
 \caption{Real space (a) and phase space (b) trajectories of 10 individually traced electrons, located initially within the time interval $t/T=0.14 \dots 0.15$ and at spatial positions $x/L=0.0633 \dots 0.07$. Discharge conditions: argon, 20 Pa, 5 cm gap, $\phi_0$ = 400 V.}
 \label{Traj_Ar_20Pa}
 \end{center}
 \end{figure}

Figure \ref{Traj_Ar_20Pa} shows real space [plot (a)] and phase space [plot (b)] trajectories of 10 individually traced electrons, located initially  within the time interval $t/T=0.14 \dots 0.15$ and at spatial positions $x/L=0.0633 \dots 0.07$, i.e. between the sheath edge and the horizontal layer of high ambipolar electric field outside the sheath during the phase of sheath expansion. Several important features are observed: when electrons collide with the expanding sheath, they are reflected and gain energy. Such a reflection can be identified in the phase-space plot (b) as a vertical rise, e.g. the blue line at $x/L \approx 0.05$. The absolute value of the electron velocity in $x$ direction after this collision with the expanding sheath is higher compared to its velocity before the collision. Typically, an electron gains about $5 \times 10^5$ m/s due to a collision with the expanding sheath. This is in good agreement with the Hard Wall Model, which predicts each electron to gain twice the sheath expansion velocity. Here, the maximum sheath expansion velocity is about $2.4 \times 10^5$ m/s. Electrons can also collide with the expanding sheaths multiple times during one sheath expansion phase, if they are reflected back towards the sheath edge by a collision. The velocity gain caused by classical sheath expansion heating is not sufficient to cause ionization and, therefore, cannot explain the observed ionization maxima. These maxima can only be explained, when considering a synergistic combination of different heating events, i.e. an individual electron must collide with the expanding sheath edge multiple times and/or be accelerated by the ambipolar electric field outside the sheath. We clearly observe that electrons gain about $7 \times 10^5$ m/s when propagating through the horizontal layer of high ambipolar field outside the sheath. Most of these electrons have not collided with the expanding sheath before. Those electrons are marked in plot (b) of figure \ref{Traj_Ar_20Pa} and leave the ambipolar layer at velocities in $x$-direction of about $1.1 \times 10^6$ m/s. This is not enough to cause ionization either. The electrons with the highest velocity in $x$ direction are those, which experienced both heating mechanisms, i.e. sheath expansion and ambipolar electron heating. After being accelerated by the expanding sheath these electrons pass the horizontal layer of high ambipolar electric field and gain about $6 \times 10^5$ m/s (blue trajectory in figure \ref{Traj_Ar_20Pa}). Such electrons leave the ambipolar layer with velocities in $x$ direction above $1.7 \times 10^6$ m/s. Assuming the velocity components in $y$ and $z$ direction to be equivalent to the thermal energy such electrons leave the ambipolar layer with total velocities, $v$, above $2.3 \times 10^6$ m/s. This is sufficient to cause ionization in argon (ionization threshold of 15.6 eV). Indeed we find most ionization events to be caused by electrons that "collided" with the expanding sheath multiple times and are accelerated by the ambipolar electric field. Electrons heated by the expanding sheath subsequently cross the region of high ambipolar electric field outside the sheath. Thus, both heating mechanisms are in phase (see figure \ref{Efield_Analysis_Ar}).

This synergistic heating of electrons by the expanding sheath and the ambipolar electric field outside the sheath explains how electrons can gain enough energy to cause ionization and provides a kinetic understanding of how CCRF plasmas operated in the $\alpha$-mode are generated. This picture is in agreement with experimental investigations of stochastic heating in such plasmas \cite{Schulze_Stoch}. We believe that these results also explain why a stronger time modulation of the electron temperature is required using the Hard Wall Model compared to fluid-kinetic models of electron heating to obtain comparable heating rates \cite{LafleurHeat1}. This might be caused by the fact that the Hard Wall Model only describes sheath expansion heating, but fluid-kinetic models include sheath expansion and ambipolar electron heating. In order to obtain the same heating rate from these two models the sheath expansion heating must be enhanced in the Hard Wall Model by a stronger time modulation of the electron mean energy. Then the same heating rate can be obtained from the Hard Wall Model compared to kinetic-fluid models that use a weaker time modulation of $T_e$, but include the combination of both heating mechanisms.



 \begin{figure}[ht!]
 \begin{center}
 \includegraphics[width=0.62\textwidth]{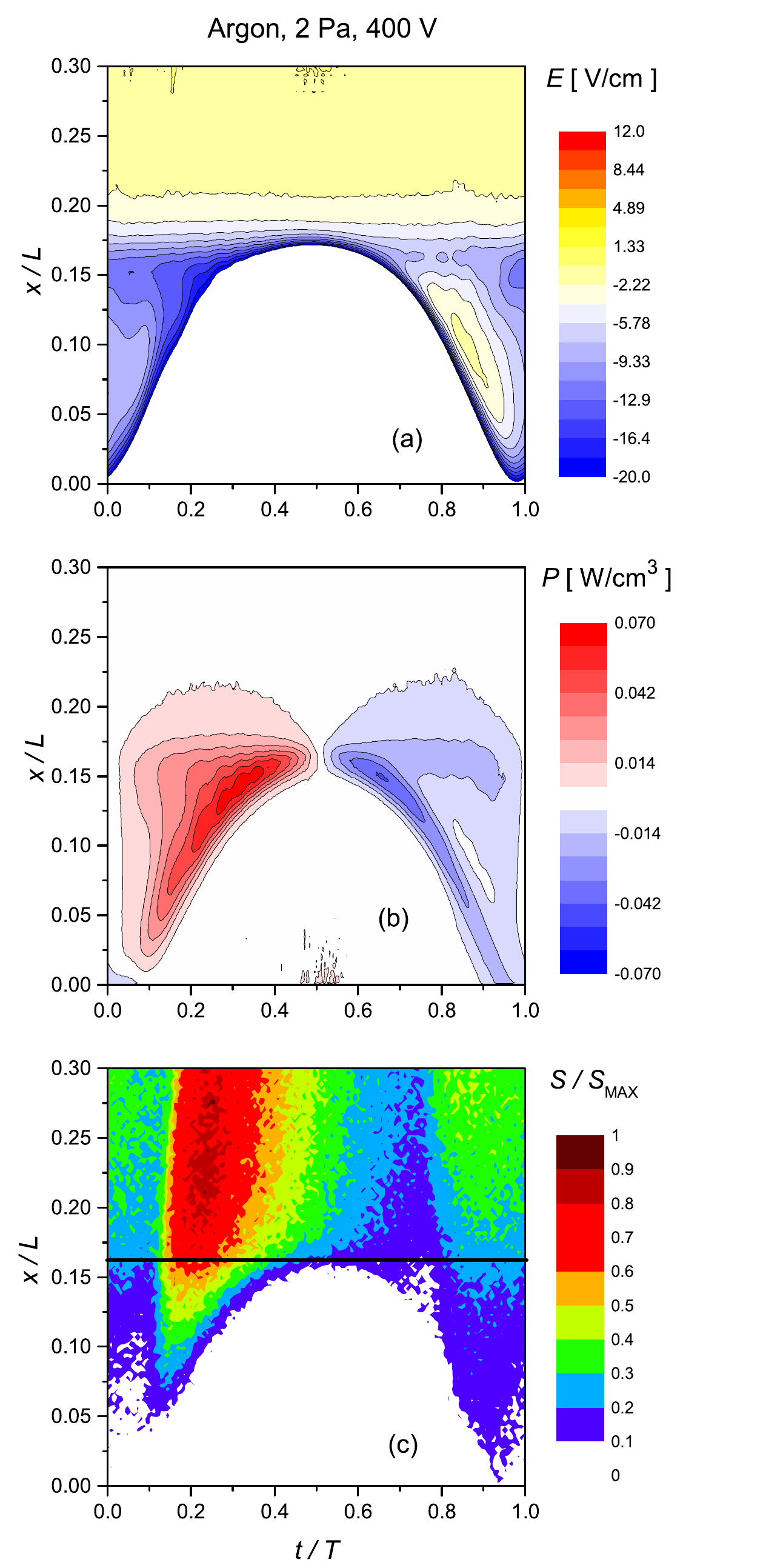} 
 \caption{Spatio-temporal plots of the electric field (a), electron heating rate, $P$ (b), and normalized ionization rate, $S/S_{MAX}$ (c). Discharge conditions: argon, 2 Pa, 5 cm gap, $\phi_0$ = 400 V. Only the spatial region close to the bottom powered electrode is shown. Time is normalized by, $T$, the length of RF period. $S_{\rm MAX}$ = 5$ \times 10^{20}$ m$^{-3}$s$^{-1}$.}
 \label{Ar2Pa}
 \end{center}
 \end{figure}

 \begin{figure}[ht!]
 \begin{center}
 \includegraphics[width=1\textwidth]{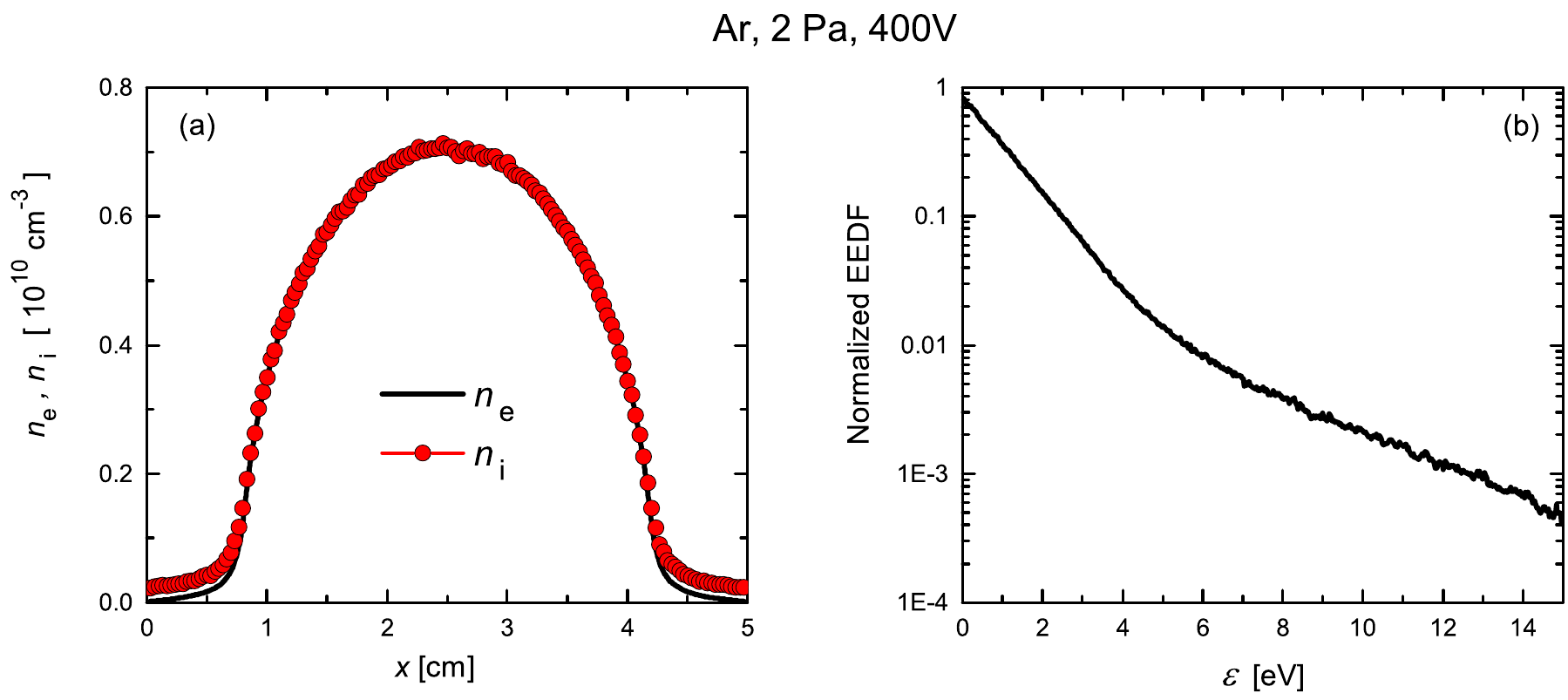} 
 \caption{Time averaged profiles of the electron and ion density as well as the time averaged normalized EEDF in the discharge center in argon at 2 Pa, 5 cm gap, and $\phi_0$ = 400 V obtained from the simulation.}
 \label{DensEEDF2}
 \end{center}
 \end{figure}

\begin{figure}[ht!]
 \begin{center}
 \includegraphics[width=0.7\textwidth]{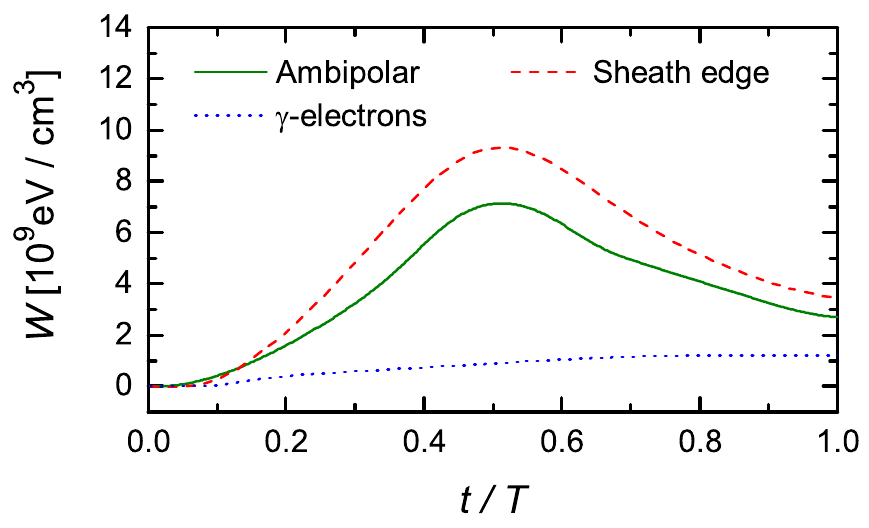} 
 \caption{Energy transferred to electrons since the beginning of the RF period, $W$, as a function of $t/T$, within one RF period in three different spatial regions: ambipolar region ($x/L=0.16$, green solid line), sheath edge ($s(t)-0.1$ mm$\leq x \leq s(t)+0.1$ mm, red dashed line), and deep inside the sheath ($x\leq s(t)-1$ mm, blue dotted line). Discharge conditions: argon, 2 Pa, 5 cm gap, $\phi_0$ = 400 V.}
 \label{W1_Ar_2Pa}
 \end{center}
 \end{figure}

\begin{figure}[ht!]
 \begin{center}
 \includegraphics[width=1\textwidth]{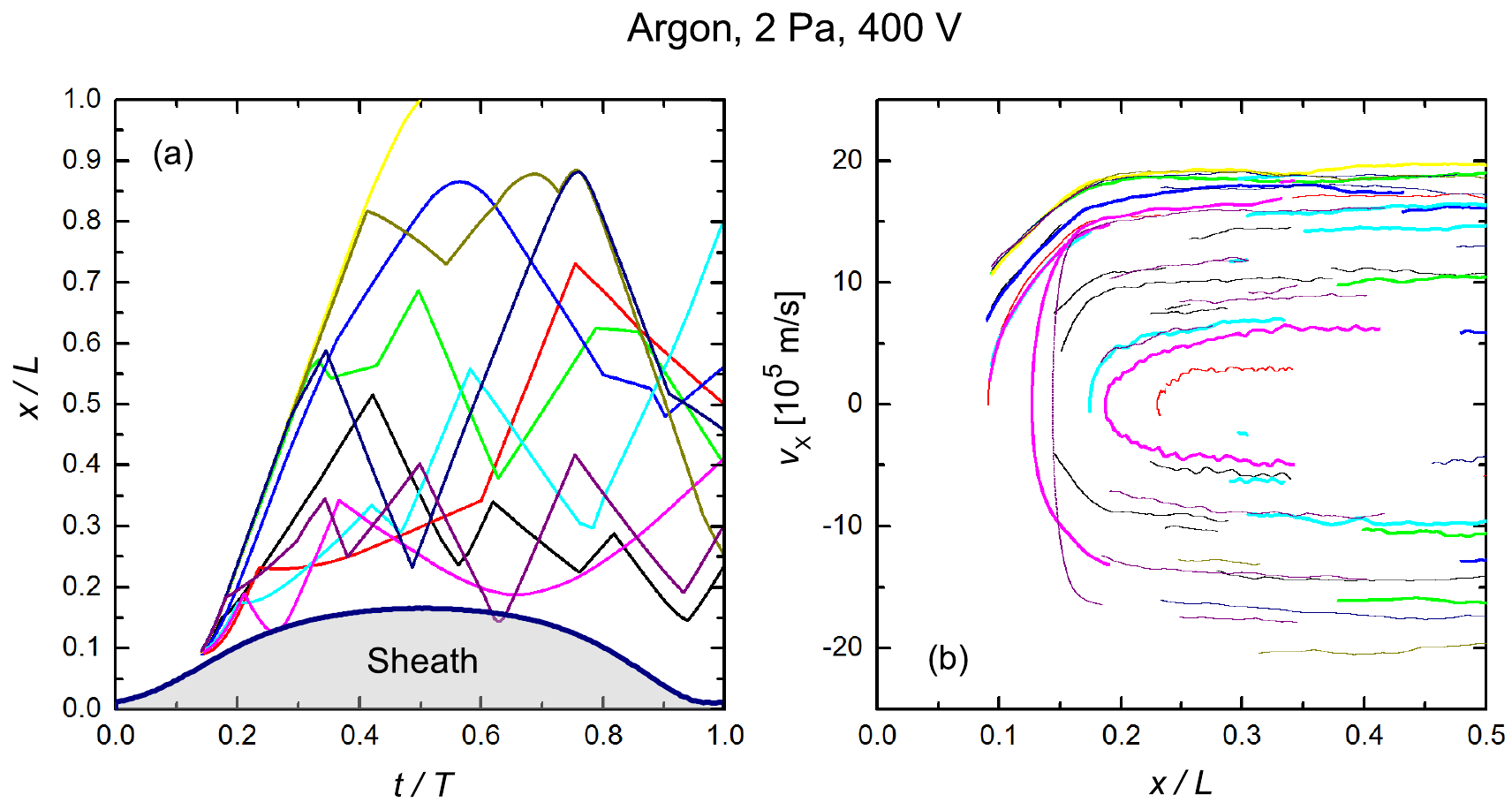} 
 \caption{Real space (a) and phase space (b) trajectories of 10 individually traced electrons, initially located within the domains $t/T=0.14 \dots 0.15$ and $x/L=0.09 \dots 0.095$. Discharge conditions: argon, 2 Pa, 5 cm electrode gap, $\phi_0 = 400$ V.}
 \label{Traj_Ar_2Pa}
 \end{center}
 \end{figure}

Similar results are found in argon at 2 Pa and $\phi_0 = 400$ V (see figs. \ref{Ar2Pa} - \ref{Traj_Ar_2Pa}). Under these conditions the sheaths are bigger due to the lower ion density and the horizontal layer of ambipolar electric field is located at $x/L \approx 0.16$. Again, there is a significant asymmetry between the expansion and collapse phase of the sheath adjacent to the powered electrode, i.e. the ambipolar field is lower during sheath collapse. This asymmetry is stronger at 2 Pa compared to 20 Pa in Argon due to the presence of a weak field reversal at 2 Pa. This effect, which will be discussed in more detail in the following section, results in a more efficient ambipolar electron heating on time average at low pressures. This asymmetry leads to less cooling during sheath collapse than heating during sheath expansion along this horizontal ``ambipolar region''. At this low pressure in argon, the bulk density is lower compared to the high pressure case and the EEDF shows an enhanced high energy tail, i.e. it is markedly non-Maxwellian (see figure \ref{DensEEDF2}). Figure \ref{W1_Ar_2Pa} shows that the ambipolar electron heating and sheath expansion heating are of similar importance under these conditions.

Figure \ref{Traj_Ar_2Pa} shows real space [plot (a)] and phase space [plot (b)] trajectories of 10 individually traced electrons, located initially between the sheath edge and the horizontal layer of ambipolar field (within the time interval $t/T=0.14 \dots 0.15$ and at spatial positions $x/L=0.09 \dots 0.095$). Under these conditions much fewer collisions are observed compared to the 20 Pa scenario. However, the same heating mechanisms are found. Electrons are accelerated to high enough energies to cause ionization by the combination of sheath expansion and ambipolar heating. When passing the horizontal layer of high ambipolar electric field outside the sheath, electrons gain about $7 \times 10^5$ m/s. 

\newpage

\subsection{Helium plasmas}

 \begin{figure}[ht!]
 \begin{center}
 \includegraphics[width=0.65\textwidth]{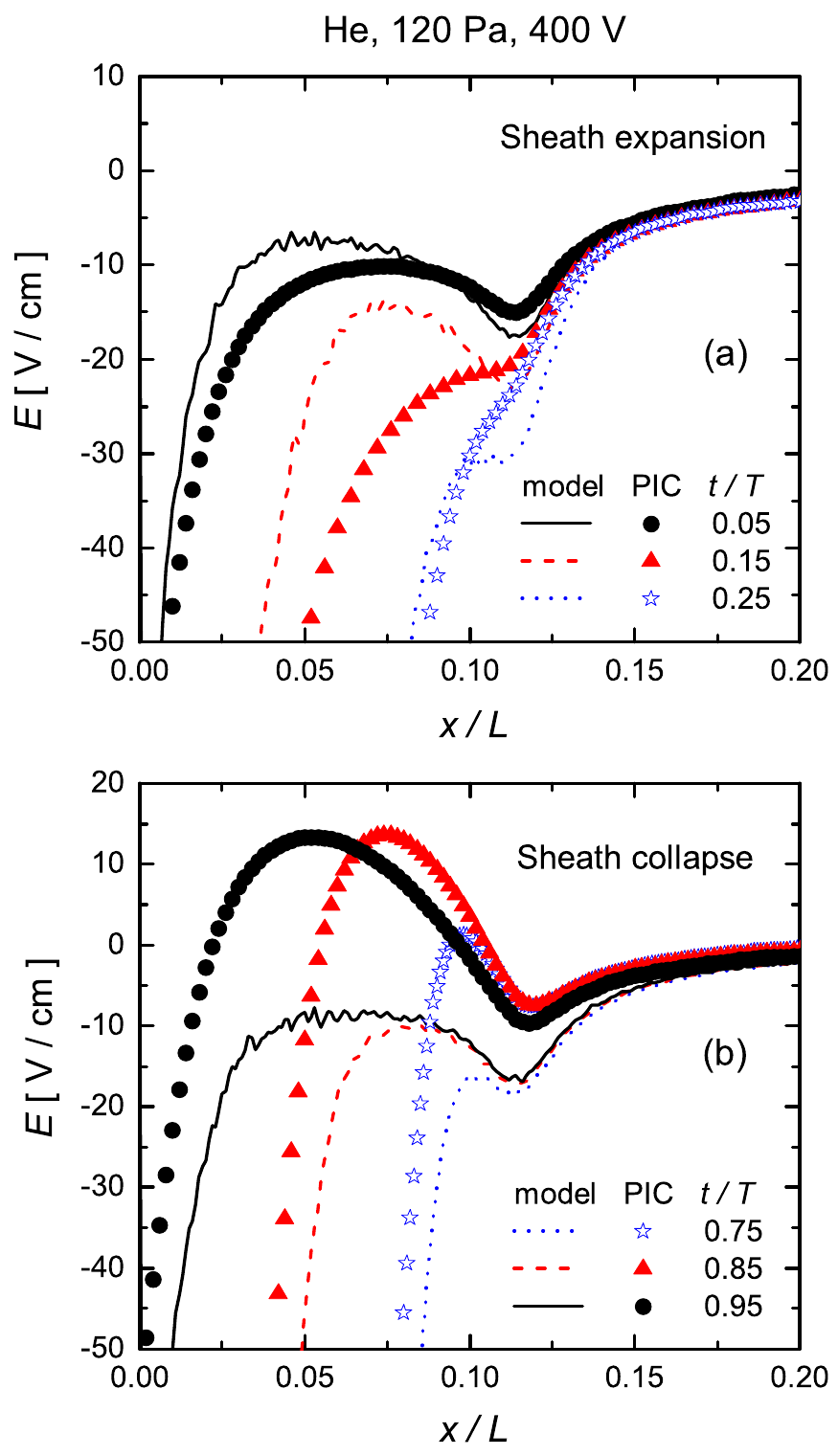} 
 \caption{Axial electric field profiles close to the powered electrode resulting from the simulation and the analytical model [equation (\ref{term4})] at 3 different times during sheath expansion (a) and collapse (b). Discharge conditions: helium, 120 Pa, 5 cm gap, $\phi_0$ = 400 V.}
 \label{Field_He_120Pa_1}
 \end{center}
 \end{figure}

In this section, we focus on a detailed analysis of a helium discharge operated at 120 Pa. The spatio-temporal distributions of the electric field, electron heating rate, and ionization rate are shown in the right column of figure \ref{overview}. These conditions were chosen, since they result in a similar maximum sheath width and expansion velocity compared to the argon case at 20 Pa, but include a strong electric field reversal during the phase of sheath collapse due to the presence of helium at high pressure \cite{SchulzeFieldRev}. This field reversal is found to reduce the electron cooling due to electron interactions with the collapsing sheath and the ambipolar electric field outside the sheaths during its collapse.

Figure \ref{Field_He_120Pa_1} shows axial electric field profiles close to the powered electrode resulting from the simulation and equation (\ref{term4}), i.e. only the fourth term of equation (\ref{EField}), at 3 different times during the phase of sheath expansion [plot (a)] and collapse [plot (b)], respectively. Similarly to argon plasmas a horizontal layer of high electric field is observed outside the sheath at $x/L \approx 0.12$ in the simulation. This electric field is again time modulated. Compared to the argon plasmas, its time modulation is much stronger, i.e. the local extremum of the electric field outside the sheath is much stronger during the phase of sheath expansion compared to the collapse. This is caused by the presence of a field reversal during sheath collapse in helium [see fig. \ref{overview} (b)], which reduces the absolute value of the electric field at $x/L \approx 0.12$. 

 \begin{figure}[ht!]
 \begin{center}
 \includegraphics[width=0.65\textwidth]{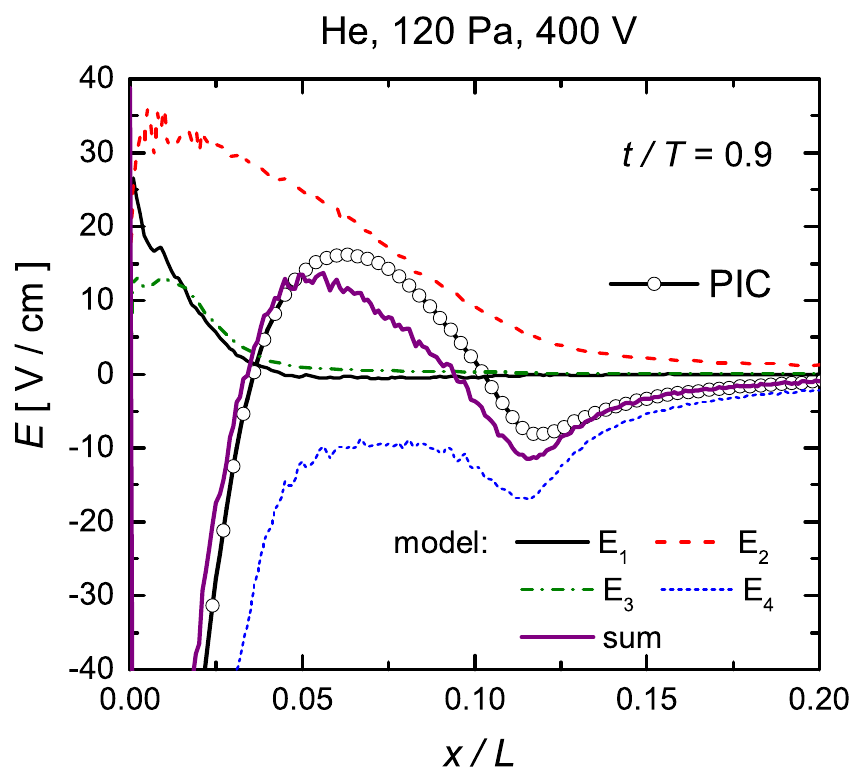} 
 \caption{Axial electric field profiles close to the powered electrode resulting from the simulation, each term of eq. (\ref{EField}), and the sum of all terms at $t/T = 0.9$ during sheath collapse. Discharge conditions: helium, 120 Pa, 5 cm gap, $\phi_0$ = 400 V.}
\label{Field_He_120Pa_b}
 \end{center}
 \end{figure}

 \begin{figure}[ht!]
 \begin{center}
 \includegraphics[width=0.7\textwidth]{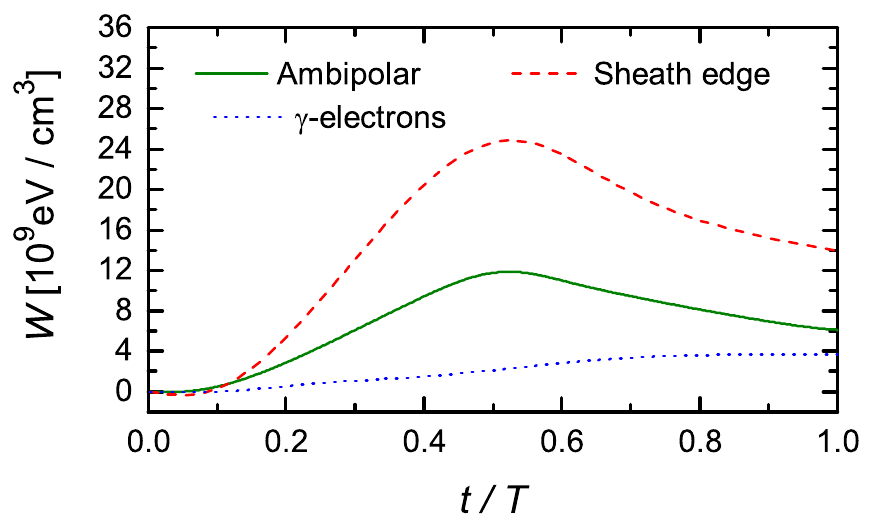} 
 \caption{Energy transferred to electrons from the beginning of the RF period, $W$, as a function of $t/T$ within one RF period in three different spatial regions: ambipolar field ($x/L=0.12$, green solid line), sheath edge ($s(t)-0.1$ mm$\leq x \leq s(t)+0.1$ mm, red dashed line), and deep inside the sheath ($x\leq s(t)-1$ mm, dotted blue line). Discharge conditions: helium, 120 Pa, 5 cm gap, $\phi_0$ = 400 V.}
 \label{HeW1}
 \end{center}
 \end{figure}

Due to the presence of the field reversal the electric field profile obtained from the simulation cannot be reproduced by equation (\ref{term4}) during sheath collapse (see fig. \ref{Field_He_120Pa_1}), since this term only describes the ambipolar electric field. The field reversal is caused by collisions of electrons with the neutral background gas and is, therefore, described by equation (\ref{term2}) (see figure \ref{Field_He_120Pa_b}). We find that equations (\ref{term1}) and (\ref{term3}), i.e. the first and third term of equation (\ref{EField}) are still negligible. Due to these collisions electrons cannot compensate the positive ion flux to the electrode during sheath collapse without the presence of a reversed field, that accelerates them towards the electrode. The second term, i.e. the collisionally induced field reversal, effectively reduces the local extremum of the electric field outside the sheath during sheath collapse.

The presence of the field reversal during sheath collapse, therefore, reduces the electron cooling during sheath collapse in helium. Its presence results in more heating within the layer of high electric field outside the sheath on time average due to a lower local electric field during sheath collapse compared to the sheath expansion phase. It also causes additional electron heating in close vicinity to the collapsing sheath edge such as shown in figure \ref{overview} (d) and, therefore, leads to more electron heating at the sheath edge on time average. Figure \ref{HeW1} illustrates these effects by showing a weaker decay of $W$ during the second half of the RF period (sheath collapse at the powered electrode) for both the ambipolar and the sheath heating. Under these conditions the sheath expansion is about twice as efficient as the ambipolar electron heating. 

\section{Conclusions}

We investigated the electron heating dynamics in electropositive argon and helium capacitively coupled RF discharges driven at 13.56 MHz by PIC/MCC simulations and an analytical model to calculate the electric field space and time resolved within the RF period. The discharges were found to be operated in the $\alpha$-mode, where ionization occurs primarily during the phase of sheath expansion at each electrode. The prevailing understanding of the heating of electrons to cause this ionization is purely based on classical sheath expansion heating, i.e. electrons are accelerated by single direct interactions with the expanding sheath edge. 

As a result of our studies we have proposed two points, which refine this picture and models of CCRF discharges:
\begin{itemize}
\item{Most of the electrons, that ionize, were found to experience two or more reflections from the sheath edge during a single sheath expansion phase,}
\item{An ambipolar electric field situated at a position of the maximum sheath edge results in additional heating on time average.}
\end{itemize}

The synergistic combination of these effects accelerates electrons to high enough energies to cause ionization. The mechanism was illustrated by tracing individual electrons close to the powered electrode during the phase of sheath expansion. This ambipolar electron heating represents a fundamental mechanism of electron heating in CCRF plasmas and is essential to understand the generation of such RF plasmas. The energy transferred to electrons within one RF period by ambipolar electron heating was found to be comparable to sheath expansion heating under all conditions investigated here.

The ambipolar electric field is time modulated within the RF period, i.e. it is stronger during the phase of sheath expansion and weaker during sheath collapse. Based on the analytical model we demonstrated that this time modulation of the electric field is caused by a time modulation of the electron mean energy, which in turn is caused by the presence of sheath expansion heating only during one half of the RF period at a given electrode. This modulation of the electric field results in an important asymmetry between the phases of sheath expansion and collapse. It leads to different axial electric field profiles during both phases. Asymmetric electric field profiles during sheath expansion and collapse have been observed experimentally before \cite{IOPConfBeams,Schulze_Stoch,UCZ_EField}, but have never been explained until now. Under our conditions the asymmetry results in more electron heating than cooling within the region of high ambipolar field outside the sheath on time average. It is more pronounced in helium compared to argon due to the presence of a field reversal during sheath collapse under the conditions investigated here. This field reversal reduces the absolute value of the electric field, where the ambipolar field is maximum, and results in less cooling during sheath collapse and, consequently, in more heating on time average.

The mechanism of ambipolar electron heating is the result of the presence of an ion density gradient, a time modulated electron mean energy, and an electric field reversal under distinct discharge conditions. We conclude that both mechanisms, i.e. ambipolar electron heating and multiple interactions between electrons and the expanding sheath edge, have to be included in models of electron heating in CCRF plasmas. 

\ack{This work has been supported by the Hungarian Scientific Research Fund through the grant OTKA K-105476.}

\bibliography{manuscript}

\end{document}